# A Survey on Reputation and Trust-Based Systems for Wireless Communication Networks

**Jaydip Sen**
Embedded Systems Research Group, Tata Consultancy Services
EPIP Industrial Estate, Whitefield Road, Bangalore-560066, INDIA
Emails: Jaydip.Sen@tcs.com

*Abstract*: Traditional approach of providing network security has been to borrow tools and mechanisms from cryptography. However, the conventional view of security based on cryptography alone is not sufficient for the defending against unique and novel types of misbehavior exhibited by nodes encountered in wireless communication networks. Reputation-based frameworks where nodes maintain reputation of other nodes and use it to evaluate their trustworthiness are deployed to provide scalable, diverse and a generalized approach for countering different types of misbehavior resulting form malicious and selfish nodes in these networks. In this paper, we present a comprehensive discussion on reputation and trust-based systems for wireless communication networks. Different classes of reputation system are described along with their unique characteristics and working principles. A number of currently used reputation systems are critically reviewed and compared with respect to their effectiveness and efficiency of performance. Some open problems in the area of reputation and trust-based system within the domain of wireless communication networks are also discussed.

## 1. Introduction

Reputation and trust are two very useful tools that are used to facilitate decision making in diverse fields of commerce and trade. In simple terms, reputation is the opinion of one entity about another. Essentially it signifies the trustworthiness of an entity [13]. Trust, on the other hand, is the expectation of one entity about the actions of another [12]. For over three decades, formal studies have been done on how reputation and trust can affect decision-making abilities under uncertain situations. However, it is only recently that the concepts reputation and trust have been adapted to wireless communication networks, as these concepts can effectively resolve many problems in these networks.

Two types of wireless networks viz. Mobile Ad hoc Networks (MANETs) and Wireless Sensor Networks (WSNs) have undergone tremendous technological advances over the last few years. This rapid development brings with it the associated risk of newer threats and challenges and the responsibility of ensuring safety, security, and integrity of information communication over these networks. MANETs, due to complete autonomy of the member nodes, and lack of any centralized infrastructure are particularly vulnerable to different types of attacks and security threats [37]. Moreover, because every node has a resource constraint, there is an incentive for each node to be programmed to selfishly guard its resources. This leads to manifestation of selfish behavior of every node that is harmful to the network as a whole. WSNs, on the other hand, involve some unique problems due to their usual operations in unattended and hostile areas. Since sensor networks are deployed with thousands of sensors for monitoring even a small area, it becomes imperative to produce sensors at very low costs. This invariably makes it impossible to produce tamper-resistant sensors. It is also very easy for an adversary to physically capture a sensor node and bypass its limited cryptographic security. The adversary can reprogram the captured node in such a way that it starts causing extreme damage to the system.

These problems can be somewhat resolved by incorporating reputation and trust-based systems in MANETs and WSNs. The nodes thus make reputation and trust guided decisions, for example in choosing relay nodes for forwarding packets for other nodes, or for accepting location information from beacon nodes [13]. This not only provides MANETs and WSNs with the capability of informed decision making, but also provides them with security against any internal attacks when cryptographic security might have been compromised. The system that discovers, records, and utilizes reputation to form trust, and uses trust to influence its behavior is referred to as a reputation and trust-based system. This paper is an effort to provide the reader with a complete understanding of reputation and trust-based systems from the wireless communication perspective.

The rest of the paper is organized as follows. Section 2 gives a brief background of reputation and trust from social perspective. Section 3 introduces MANETs and WSNs and discusses various types of misbehavior that may be exhibited by nodes in these networks and the effects of such misbehavior on the network performance. Section 4 discusses various characteristics of trust metric and different classes of reputation and trust-based systems for wireless communication networks along with their desirable properties. Section 5 presents a detailed discussion on various important design issues of reputation and trust-based systems in terms of how the reputation information are gathered, processed, modeled, disseminated, and responses actions taken. Section 6, makes a critical review of some of the important reputation and trust-based systems for MANETs and WSNs that have been proposed in the literature in the recent past. The strengths and weaknesses of these systems are discussed and their effectiveness compared with respect to various parameters. Section 7 highlights some of the open issues in the domain of reputation and trust from wireless communication network perspective. Section 8 concludes the paper.

## 2. Social Perspective of Trust

This section briefly introduces the concept of trust and uncertainty from social perspective as trust and uncertainty play a very crucial role in market dynamics, allowing us to model both consumer and seller behavior. Trust and uncertainty are defined and various trust antecedents are discussed in brief.

### 2.1 Trust and Uncertainty

Trust has been widely recognized as an important factor affecting consumer behavior, especially in the context of e-commerce where uncertainty exists. In fact, the concept of trust becomes even more important in situations where there are uncertainties. Research on transaction economic has already established the fact that uncertainty increases transaction cost and decreases acceptance of online shopping [1]. Trust contributes to the success of e-commerce in a long way by helping consumers handle uncertain situations. Trust is complex and multi-dimensional in nature. Many research groups have explored on various antecedents of trust. Major antecedents of trust include calculus-based trust, knowledge-based trust, and institution-based trust [2][3][4]. Uncertainty originates from two sources: information asymmetry and opportunism. The former refers to the fact that either party may not have access to all the information it needs to carry out the transaction. The latter indicates that the two parties involved in a transaction have different goals, and both try to behave opportunistically to satisfy their own interests.

### 2.2 Trust Antecedents

There exist primarily three different types of trust antecedents: (i) Calculus-based, (ii) Knowledge-based, and (iii) Institution-based. In economic transactions, parties develop trust in a calculative manner [5]. To make a calculus-based trust choice, one party rationally calculates the costs and benefits of other party's cheating or cooperating in the transaction. Calculus-based trust develops because of creditable information regarding the intention or competence of the trustee. The creditable information, such as reputation and certification can signal that the trustee's claims of trustworthiness are true [6]. Trust can also develop as a result of the aggregation of trust related knowledge by the parties involved in a transaction [7]. One of the knowledge-based antecedents for trust is familiarity. It reduces environmental uncertainty by imposing a structure. In institution-based trust one believes that the necessary impersonal structures are in place to enable one to act in anticipation of a successful future endeavor [8]. The concept of institution-based trust evolves from sociology, which deals with the structures (e.g. legal protections) that make an environment trustworthy [2]. Research has found that institution-based antecedents positively affect trust in e-vendors [4].

## 3. Trust in Wireless Communication Network

This section discusses how the concept of trust can augment security in a wireless networking environment. In the domain of wireless communication also we find information asymmetry and opportunism. Just as in e-commerce, nodes in MANETs and WSNs have no way of gathering information about nodes situated outside their radio range, and have a great deal of uncertainty. Also in systems having asymmetrical designs, some nodes may be more powerful than others and may have access to information that others do not have.

The following subsections give a brief background on MANETs and WSNs, the challenges faced in designing a reputation and trust-based systems for these networks, different types of misbehavior a malicious node can show in these networks and the effects of these misbehaving nodes on the network performance.

### 3.1 Wireless Communication Networks

A MANET is a self-configuring system of mobile nodes connected by wireless links. The nodes are free to move randomly that leads to rapid change in the topology of the network. The network lacks any centralized infrastructure, and therefore all network activities are carried out by the nodes themselves. Every node acts both as an end-system as well as a relay node that forwards packets for other nodes. Since MANETs do not require any fixed infrastructure, they are highly preferred for quickly setting up networks for connecting a set of mobile devices in emergency situations like rescue operations, disaster relief efforts or in other military operations. MANETs can either be managed by an organization that enforces access control or they can be open to any participant that is located close enough. The later scenario poses greater security threats. In MANETs, nodes are autonomous and do not have any common interest. It may seem to be advantageous for a node not to cooperate with other nodes in the network and behave selfishly. Hence, the nodes need some sort of incentive and motivation so that they start cooperating each other. The non-cooperative behavior of a node may be due to selfish intention, for example to save power, or malicious intention, for example to launch denial-of-service attacks.

A WSN is a network of hundreds and thousands of small, low-power, low-cost devices called sensors. The core application of WSNs is to detect and report events. WSNs have found critical applications in military and civilian domain, including robotic landmine detection, battlefield surveillance, environmental monitoring, wildfire detection and traffic regulation. They have invaluable contributions in life saving operations, be it the life of a soldier in the battlefield, or a civilian's life in areas of high chances of natural calamities. In WSNs all the sensors belong to a single group or entity and work towards the same goal, unlike in MANETs. An individual senor has little value of its own unless it works in cooperation with other sensors. Hence, there is an inherent motivation for nodes in WSNs to be cooperative, and so incentive is less of a concern. Since WSNs are often deployed in unattended territories that can often be hostile, they are vulnerable to physical capture by enemies. An obvious solution to this problem is to make the senor nodes tamper-proof. However, this makes the sensor nodes prohibitively expensive to manufacture. Since many nodes are often required to cover an area, each node must be cheap to make the use of the network economically feasible. As tamper-proofing the nodes is not a viable solution, an adversary can modify the sensors in such a way that they start misbehaving and disrupt communication in the network. It may be even possible for the adversary to break the cryptographic security of the captured node and launch attacks from within the network as an insider. Even though cryptography can provide integrity, confidentiality, and authentication, it cannot defend against an insider attack. This necessitates a security mechanism inside a WSN that can cope with insider attacks.

### 3.2 Misbehavior of Nodes

The lack of infrastructure and organizational environment of MANETs and WSNs make these networks particularly vulnerable to different types of attacks. Without proper countermeasures, it is possible to gain various advantages by malicious behavior: better service than cooperating nodes, monetary benefits by exploiting incentive measures or trading confidential information, saving power by selfish behavior, preventing someone else from getting proper service, extracting data to get confidential information, and so on. Even if the misbehavior is not intentional, as in the case of a faulty node, the effects can be detrimental to the performance of a network. The non-cooperative behavior of a node in a MANET as identified in [9], is mainly caused by two types of misbehavior: selfish behavior e.g., nodes that want to save power, CPU cycles, and memory, and malicious behavior which are not primarily concerned with power or any other savings but interested in attacking and damaging the network. Karlof and Wagner [38] have identified various types of security threats in a WSN due to malicious nodes and proposed some countermeasures of them. When the misbehavior of a node manifests as selfishness, the system can still cope with it since this misbehavior can always be predicted. A selfish node will always behave in a way that maximizes its benefits, and as such, incentive can be used to ensure that cooperation is always the most beneficial option. However, when the misbehavior manifests as maliciousness, it is hard for the system to cope with it, since a malicious node always attempts to maximize the damage caused to the system for its own benefit. As such, the only method of dealing with such a node is detection and isolation from the network. Malicious misbehavior in packet forwarding can be generally divided into two types: forwarding misbehavior and routing misbehavior. Some common examples of forwarding misbehavior are packet dropping, modification, fabrication, timing attacks, and silent route change. Packet dropping, modification, and fabrication are self-explanatory. Timing misbehavior is an attack in which a malicious node delays packet forwarding to ensure that the time-to-live (TTL) of the packets are expired, so that it is not immediately understood by other nodes. A silent route change is an attack in which a malicious node forwards a packet through a different route than it was intended to go through. Routing misbehavior may include route salvaging, dropping of error messages, fabrication of error messages, unusually frequent route updates, silent route changes, and sleep deprivation. In route salvaging attack, the malicious node reroutes packets to avoid a broken link, although no error actually has taken place. In silent route change attack, a malicious node tampers with the message header of either control or data packets. In sleep deprivation attack, a malicious node sends excessive number of packets to another node so as to consume computation and memory resources of the latter. There exist three other types of routing misbehavior: blackhole, grayhole, and wormhole. A blackhole attack is one in which a malicious node claims to have the shortest path but when asked to forward the packets, it drops them. In a grayhole attack, which is a variation of the blackhole attack, the malicious node selectively drops some packets. A wormhole attack, also known as tunneling, is an attack in which the malicious node sends packets from one part of the network to another part of the network, where they are replayed.

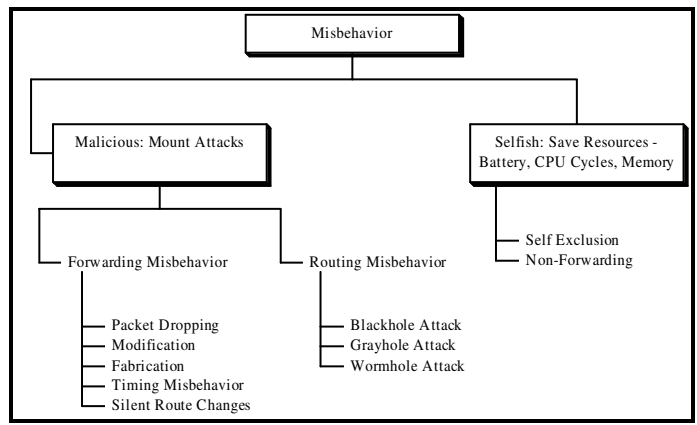

**Figure 1: Nodes' Misbehavior in MANETs and WSNs**

The selfish behavior of a node can be generally classified as either self-exclusion or non-forwarding. The self-exclusion misbehavior is one in which a selfish node does not participate when a route discovery protocol is executed. This ensures that the node is excluded from the routing list of other nodes. This benefits a selfish node by helping it save its power, as it is not required to forward packets for other nodes. A reputation model is an effective way to thwart the intentions of such selfish nodes. Since a node does not forward packets for other nodes in the networks, it is denied any cooperation by other nodes. So, it is in the best interest of a selfish node to be cooperative. On the other hand, the non-forwarding misbehavior is one in which a selfish node fully participates in route discovery phase but refuses to forward the packets for other nodes at a later time. This selfish behavior of a node is functionally indistinguishable from a malicious packet dropping attack.

Since reputation-based systems can cope with any kind of observable misbehavior, they are useful in protecting a system. Reputation and trust-based systems enable nodes to make informed decisions on prospective transaction partners. Researchers have been steadily making efforts to successfully model WSNs and MANETs as reputation and trust-based systems. Adapting reputation and trust-based systems to WSNs presents greater challenges than MANETs and Peer-to-Peer (P2P) systems due to their energy constraints. CORE [10], CONFIDANT [11], RFSN [12], DRBTS [13], KeyNote [14], and RT Framework [15] are some of the well-known systems in this area. However, RFSN and DRBTS are the only works so far focusing on WSNs. The others concentrate on MANETs and P2P networks.

### 3.3 Effects of Nodes' Misbehavior

In wireless networks without appropriate countermeasures, the effects of misbehavior have been shown by several simulations is to dramatically decrease network performance [9][11][16]. Depending on the proportion of misbehaving nodes and their specific strategies, network throughput can be severely degraded, packet loss increased, and denial-of-service experienced by honest nodes in the network. In a theoretical analysis of how much cooperation can help by increasing the probability of a successful forwarding of packets, Lamparter, Plaggemeir, and Westhoff have found that increased cooperation more than proportionately increases the performance for small networks with fairly short routes. Zhang and Lee [17] argue that prevention measures, such as encryption and authentication can be used in MANETs to reduce the success of intrusion attempts, but cannot completely eliminate them. For example, encryption and authentication cannot defend against compromised mobile nodes, which carry the private keys. No matter what types of intrusion prevention measures are deployed in the network, there are always some weak links that an adversary can exploit to break in. Intrusion detection presents a second wall of defense and it is a necessity in any high-survivability network.

## 4. Reputation and Trust-Based Systems

Use of reputation and trust-based systems for Internet, e-commerce and P2P applications has been there for over half a decade [18][19][20][21][22]. However, it is only recently that efforts have been made to model MANETs and WSNs as reputation and trust-based systems [9][11][23][24]. This section presents various characteristics of trust, the goals and properties of reputation systems and different types of reputation and trust-based systems.

### 4.1 Trust and its Characteristics

Form the perspective of wireless communication networks, Sun et al [34] have identified some characteristics of trust metric. These characteristics are as follows:

(i) Trust is a relationship established between two entities for a specific action. In particular, one entity trusts the other entity to perform an action. The first entity is called the subject, the second is called the agent.
(ii) Trust is a function of uncertainty. In particular, if the subject believes that the agent will perform the action for sure, the subject fully trust the agent to perform the action and there is no uncertainty; the subject believes that the agent will not perform the action for sure, the subject trusts the agent not to perform the action, and there is no uncertainty either; if the subject does not have any idea of whether the agent will perform the action or not, the subject does not have trust in the agent, In this case, the subject has the highest level of uncertainty.
(iii) The level of trust can be measured by a continuous real number, refereed to as the trust value. Trust value should represent uncertainty.
(iv) The subjects may have different trust values with the same agent for the same action. Trust is not necessarily symmetric. The fact that *A* trusts *B* does not necessarily mean that *B* also trust *A*, where *A* and *B* are two entities.

### 4.2 Reputation Systems- Goals and Properties

The important goals of reputation and trust-based systems for wireless communication networks have been identified in [20]. They are as follows: (i) provide information that allows nodes to distinguish between trustworthy and untrustworthy nodes in the network, (ii) encourage the nodes in the network to cooperate with each other and become trustworthy, (iii) discourage the untrustworthy nodes to participate in the network activities. We have identified two additional goals of a reputation and trust-based system from wireless communication network perspective. The first goal is to be able to cope with any kind of observable misbehavior, and the second is to minimize the damage caused by any insider attacks.

To operate effectively and efficiently, reputation and trust-based system for wireless communication networks must have three essential properties as identified in [18]. These properties are as follows: (i) the system must have long-lived entities that inspire expectations for future interactions, (ii) the system must be able to capture and distribute feedbacks about current interactions among its components and such information must also be available in future, and (iii) the system must use feedback to guide trust decisions.

**4.3 Classification of Reputation and Trust-Based Systems**

There are different perspectives from which reputation and trust-based systems can be classified. They can be categorized from the perspective in which they are initialized, the types of observation they use, the manner in which the observations are accessed, and the way the observed information is distributed in the network. These are discussed in detail below.

Most of the trust and reputation-based systems are initialized in one of the following three ways:
  i. All the nodes in the networks are initially assumed to be trustworthy. Every node trusts other nodes in the network. The reputations of the nodes decrease with every bad encounter.
  ii. Every node is considered to be untrustworthy in the system bootstrapping stage, and the nodes do not trust each other initially. The reputations of nodes with such system increase with every good encounter.
  iii. Every node in the network is considered to be neither trustworthy nor untrustworthy. All nodes start with a neutral reputation value to start with. With every good or bad behavior, the reputation value is increased or decreased respectively.

On the basis of observations they use, reputation and trust-based system can be classified into two groups: (i) systems using only first-hand information and (ii) systems using both first-hand and second-hand information. While the systems using the first-hand information rely on the direct observations or experiences encountered by the nodes, the nodes in the systems using second-hand information utilize information provided by the peers in its neighborhood. Most of the current reputation systems use both first-hand and second-hand information to update reputation. This allows the systems to make use of more information about the network in computing reputation values. There are systems that use only first-hand information. This makes the systems completely robust against rumor spreading. OCEAN [25] and Pathrater [23] are two such systems. In DRBTS [13], certain types of nodes use only second-hand information. In this system, a node does not have any first-hand information to evaluate the trustworthiness of the informers. One way to deal with this reputation is to use a simple majority principle. Reputation systems can be broadly categorized into two types depending on the manner in which different nodes access reputation information in the network. These two types are: (i) symmetric systems and (ii) asymmetric systems. In symmetric reputation systems, all nodes in the network have access to the same level of information, i.e. both first-hand and second-hand information. In asymmetric systems, on the other hand, all nodes do not have access to the same amount of information. For example, in DRBTS [13], sensor nodes do not have first-hand information. Thus, in the decision making process, the sensor nodes are at a disadvantageous situation due to lack of availability of information.

On the basis of the manner in which reputation is distributed in the network reputation systems can be categorized into two groups: (i) centralized and (ii) distributed. In centralized systems, one central entity maintains the reputations of all nodes in the network. This central entity can be a source of security vulnerability and performance bottleneck in the system. Examples of this type are eBay and Yahoo auctions. In distributed systems, each node maintains reputation information of all the nodes about which it is interested. In such systems, maintaining consistency in reputation values maintained in different nodes may be a major challenge. In a distributed system each node may maintain reputation of the nodes that are within its communication range, or may maintain reputation information of all the nodes in the network. In sensor network applications, every node maintains reputation information only for its neighbors. This reduces the memory overhead for reputation information maintenance. However, for networks with high mobility, maintenance of reputation for as many nodes as possible is a preferred option for every node. This ensures that a node does not get completely alienated if it moves to a new location with a changed neighborhood. This strategy, of course, involves a very large memory overhead.

Irrespective of the type of a reputation and trust-based system, its objective should be effectively detect and isolate misbehaving nodes in the network. It should be self-organized and robust against any insider attacks. The reputation computation and maintenance system should not be vulnerable to manipulation by a malicious attacker. Moreover, it should not involve much memory and communication overhead. All these criteria make designing an effective and efficient reputation system for MANETs and WSNs an extremely challenging task

**5. Issues in Reputation Systems for Wireless Communication Networks**

This section discusses various issues of reputation and trust-based systems. Several important design parameters of a reputation and trust-based systems for MANETs and WSNs and discussed in detail illustrating them with real world systems whenever appropriate.

## 5.1 Information Gathering

Information gathering is the process by which a node collects information about other nodes it is interested in. This is concerned only with first-hand information. First-hand information is gathered by a node purely on the basis of its observation and experience. However, in CONFIDANT [11], first-hand information is further classified into personal experience and direct observation. Personal experience of a node refers to the information it gathers through one-to-one interaction with its neighbors. Direct observation is the information gathered by a node by observing the interactions among its neighbors. CONFIDANT [11] is currently the only system that makes this distinction.

Most reputation and trust-based systems make use of a component called Watchdog [23] to monitor their neighborhood and gather information based on promiscuous mode of observation. Thus, first-hand information is confined to the wireless sensing range of a node. However, the watchdog system is not very effective in situations where directional antennas are deployed and spread spectrum technology is used for wireless communication. This aspect is getting lot of focus in current research activities on wireless communications.

## 5.2 Information Dissemination

There is an inherent trade-off between the efficiency in using second-hand information and robustness against false ratings. Use of second-hand information gives lot of advantages. Firstly, the reputation of the nodes builds up more quickly due to the ability of the nodes to learn from the mistakes of each other. Secondly, no information in the system goes unused. Finally, over a period of time, a consistent local view stabilizes in the system.

However, sharing information makes the system vulnerable to *false report attacks*. This vulnerability can be can be somewhat reduced by adopting a strategy of limited information sharing, i.e., sharing either only positive information or negative information.

If only positive information is shared, the system is still vulnerable to false praise attacks. With only positive information being shared, the nodes cannot share their bad experiences. This is particularly detrimental since learning from ones own experience in this scenario comes at a very high price. Also, colluding malicious nodes can extend each other's survival time through false praise reports. CORE [10] permits only positive second-hand information, which makes it vulnerable to spurious positive ratings by malicious nodes.

Sharing only negative information protects the system against the false praise attack, but it has its own drawbacks. The nodes cannot share their good experiences. More importantly, malicious nodes can launch bad-mouth attacks on benign nodes either individually or in collusion with other malicious nodes. CONFIDANT [11] makes use of negative second-hand information in order to proactively isolate misbehaving nodes. This makes the system vulnerable to spurious ratings, and false accusations. Context-aware detection [26] accepts negative second-hand information on the condition that at least four separate sources make such a claim, otherwise the node spreading the information is considered misbehaving. While this distributes the trust associated with the accusation over several nodes and thus distributes the risk, it inadvertently serves as a disincentive to share ratings and warn other nodes by spreading reputation information in the network. It is also not possible to guarantee the availability of four witnesses for an event in a sparsely populated network.

Another way of avoiding the negative consequences of information sharing is not to share any information at all. OCEAN [25] is one such model that builds reputation purely based on the individual observations of the nodes. Although such systems are completely robust against rumor spreading, they have some shortcomings. The time required to build reputation is increased dramatically, and it takes longer duration for reputation to fall, allowing malicious nodes to stay in the system and misuse the system resources.

Systems like DRBTS [13] and RFSN [12] share both positive and negative information. The negative effects of information sharing, as discussed above, can be mitigated by appropriately incorporating first-hand and second-hand information into the reputation metric. Using different weighting functions for different information is one efficient technique.

Most of the reputation and trust-based systems for MANETs and WSNs use one of the three following methods to share information among the nodes: *friends list*, *blacklist*, and *reputation table*. A friends list shares only positive information, a blacklist shares only negative information, while a reputation table shares both positive and negative information.

Information sharing involves three important issues: (i) dissemination frequency, (ii) dissemination locality, and (iii) dissemination content. These issues are briefly discussed below.

The reputation systems can be of two types on the basis of dissemination frequency they employ: (i) Proactive dissemination and (ii) Reactive dissemination. In proactive dissemination, nodes communicate reputation information during each dissemination interval. A node publishes the reputation values even if there have been no changes in the stored values in the last dissemination interval. This strategy is more suited to dense network with more activities, as the nodes have to wait till the beginning of the next dissemination interval to publish their reputation information. In reactive dissemination, nodes publish only when there is predefined amount of change to the reputation values they store or when an event of interest occurs. This method reduces communication overhead in situations where reputations of nodes do not change frequently. However, reactive dissemination may cause congestion in networks with high network activity. In both these types of information dissemination, the communication overhead can be reduced to a large extent by piggy backing the information with other network traffic. In CORE [10], the reputation information is piggybacked on the reply messages and in DRBTS [13] it is piggybacked on the location information dispatch messages.

Reputation systems may use two types of locality of dissemination of information: (i) local, and (ii) global. In local dissemination, the information is published within the neighborhood. It could be either through a local broadcast, multicast, or unicast. In DRBTS [13], the information is published in the neighborhood through a local broadcast. This enables all the beacon nodes to update their reputation table accordingly. A reputation system may also choose to unicast or multicast depending on the application domain and security requirements. In global dissemination, the information is propagated to nodes outside the radio range of the node publishing the reputation information. Global dissemination may also use either broadcast, multicast or unicast technique. For networks with higher node mobility, global dissemination is preferred as it provides nodes with a reasonable understanding of the new locations they are moving to.

There can be two types of reputation information contents that may be disseminated: (i) raw information and (ii) processed information. In case of raw information, the information published by a node is its first-hand information only. It does not reflect the final composite reputation value, as it does not take into consideration the second-hand information of other nodes in the neighborhood. In case of processed information, a node publishes the overall reputation values after computing the composite reputation score

### 5.3 Redemption and Weighting of Time

An important issue in maintaining and updating reputation is how past and current information are weighted. Different models weight them differently, each with a different rationale. CORE [10] assigns more weight to the past behavior of a node than its current behavior, so that wrong observations or rare behavior changes cannot influence the reputation rating too much. It helps benign nodes that may behave selfishly due to genuinely critical battery conditions. Nodes may also misbehave temporarily due to technical problems like link failure. CONFIDANT [11] takes the opposite approach- it discounts past behavior by assigning less weight. This ensures that a node cannot leverage on its past good performance and start misbehaving without being punished by making the system more responsive to sudden behavior changes of nodes. RFSN [12] also give more weight to recent observations than the past. This forces nodes to be cooperative at all the time. However, there is a problem in adopting the strategy of assigning higher weights to current behavior. In periods of low network activity, a benign node may get penalized. This problem can be resolved by generating network traffic in regions and periods of low network activity using mobile nodes. DRBTS [13] tackles this issue by generating network traffic through beacon nodes when a need arises. Pathrater [23], Context-aware detection [26], and OCEAN [25] do not weight ratings according to time.

Ratings are not only weighted to put emphasis on the past or the present, but also to add importance to certain kinds of observation. CONFIDANT [11] gives more weight to first-hand observations and less to reported second-hand information. CORE [10] also assigns different weights to different types of observations.

Redemption is done in case a node is wrongly identified as a misbehaving node, either because of deceptive observation, spurious ratings, or a fault in the reputation system. Redemption is also necessary when a node that was previously isolated from the network because of its misbehavior needs to be allowed to join back, because the cause of its misbehavior has been identified and resolved, e.g., a faulty node may have been repaired, a compromised node may have been recaptured by its legitimate user.

CONFIDANT [11] carries out redemption of misbehaving or misclassified nodes by reputation fading, i.e. discounting the past behavior even in the absence of testimonials and observations, and periodic reevaluation, i.e., checking from time to time whether the rating of a node is above or below the acceptable threshold. Thus a node that has been isolated from the network because of its misbehavior, always gets a chance to rejoin the after some time. Since the ratings do not get erased but only discounted, the rating of a previously misbehaving node will still be close to the threshold value and thus the reaction to a current misbehavior will be swift. This will result in faster detection and isolation of that node in case it starts misbehaving. It is thus possible for a node to redeem itself. Since the nodes in the network may differ in their opinion, it is quite likely that a node will not be excluded by all other nodes and thus it can participate partially in the network activities. This will give the node a chance to show good behavior and redeem its reputation value. Even if this is not the case and the suspect node is excluded by everyone, it can redeem itself by means of the reputation fading.

In CORE [10], a node that is isolated because of its misbehavior in the past cannot redeem itself till there is a sufficient number of new nodes arriving in the network that have no past experience with it.

OCEAN [25] relies on a timeout of reputation. The sudden lapse back into the network can pose a problem if several nodes set the timer at roughly the same time. Pathrater [23] and Context-aware detection [26] have no provision of redemption.

### 5.4 Weighting of Second-Hand Information

The schemes that use second-hand information have to administer trust of the witness, i.e., the sources of second-hand information, in order to prevent blackmailing attacks. It is thus necessary to use some means of validating the credibility of the reporting node. One method is to use a deviation test as done in [13] and [27]. If the reporting node passes the deviation test, it is treated as trustworthy and its information is incorporated to update the reputation of the reported node. However, different models choose different strategies for dealing with the second-hand information depending on the application domain and security requirements. For instance, the model presented in [12] uses Dempster-Shafer theory [28] and discounting belief principle [29] to incorporate second-hand information. However, Beta distribution is mostly used in reputation and trust-based systems. It was first used by Josang and Ismail [22]. Many researchers in the field of security in ad hoc networks have used Beta distribution in their analysis. Ganeriwal, Srivastava [12] and Bucheggar, Boudec [27] are among them. The reason for popularity of Beta distribution is its simplicity as it is indexed by only two parameters.

CONFIDANT [11] assigns weights on the second-hand information according to the trustworthiness of the source and by setting a threshold that had to be exceeded before the second-hand information is taken into account. Second-hand information had to come from more than one trusted source or several partially trusted sources, or any combination thereof, provided that trust times the number of nodes exceeds the trust threshold. This adds a vulnerability to the system where some untrustworthy nodes may be trusted. The notion of trust has been more specifically defined in the enhanced version of CONFIDANT, known as RRS [27]. In RRS, trust means consistent good performance as a witness, measured as the compatibility between first and second-hand information. This dynamic assessment allows the system to keep track of trustworthiness and to react accordingly. If the second-hand information is accepted, it will have a small influence on the reputation rating. More weight is given to the nodes' direct observations.

### 5.5 Spurious Ratings

If second-hand information is used to influence reputation, some nodes may lie and give spurious rating about others. A malicious node may be benefited by falsely accusing an honest node, as this can lead to a denial of service to the latter. A false praise can benefit a colluding malicious node. Problems related to false accusations are absent in positive reputation systems, since no negative information is maintained [19][30], however, the disseminated information could be false praise and result in a good reputation for some malicious nodes. Even if the disseminated information is correct, it may not be possible to distinguish between a misbehaving node and a new node that has just joined the network.

If second-hand information is used, an important issue is to decide whether the lying nodes should be punished in the same way as the misbehaving nodes by isolating them from the network services. If nodes are punished for their seemingly inaccurate testimonials, we may end up punishing an honest messenger. This will definitely discourage honest reporting of observed misbehavior. The testimonial accuracy is evaluated according to affinity to the belief of the requesting node along with the overall belief of the network as gathered over time. The accuracy is not measured as compared to the actual true behavior of a node, since the latter is unknown and cannot be proved beyond doubt. Even if it were possible to test a node and obtain a truthful verdict on its nature, a contradicting previous testimonial could still be accurate. Thus, instead of punishing deviating views, it is better to merely reduce their impact on public opinion. Some node is bound to be the first witness of another node's misbehavior, and thus its report will start deviating from the public opinion. Punishing this discovery would be counterproductive, as the goal is precisely to learn about the misbehaving nodes as early as possible.

### 5.6 Identity

The question of identity is of central importance to any reputation systems. Identity can be of three types: *persistent*, *unique* and *distinct*. A node cannot easily change its persistent identity. Identity persistence can be achieved by expensive pseudonyms or by a specific security module. Identity persistence is desirable for a reputation system to enable it to gather the behavior history of a node. An identity is unique, if no other node can impersonate the node by using its identity. This can be achieved by cryptographically generated unique identifiers, as proposed by Montenegro and Castelluccia [31]. This property is needed to ensure that the observed behavior was indeed that of the node observed. The requirement of distinct identities is the target of the so-called Sybil attack, as analyzed by Douceur [32], where a node generates several identities for itself to be used at the same time. This property is not of much concern to the reputation system, since those identities that exhibit misbehavior will be excluded while other identities stemming from the same node will remain in the network as long as they behave well. The Sybil attack can, however, influence public opinion, by having its rating considered more than once. A mechanism to defend Sybil attack has been proposed in [33].

### 5.7 Detection

Reputation systems require a tangible object of observation that can be identified as either good or bad. In online auction or trading systems, this is a sale transaction with established and measurable criteria such as delivery or payment delay. In case of reputation systems for MANETs, the analogy of a transaction is not straightforward due to the limited observability and detectability of a mobile node. In order to detect misbehavior, nodes promiscuously overhear the communications of their neighbors. The component used for this kind of observation is called Watchdog [23], Monitor [11] or NeighborWatch [25].

The function mostly used to implement the detection component in reputation systems is *passive acknowledgement* [35], where nodes register whether their next hop neighbor on a given route has attempted to forward a packet. Assuming bi-directional links, a node can listen to the transmission of another node that is within its radio range. If within a given time window, a node hears a retransmission of a packet by the next hop neighbor, it has sent packet previously, the behavior is judged to be good. This does not necessarily mean that the packet has been transmitted successfully, since the observing node cannot see what goes on outside its radio range, e.g., there could be a collision on the far side of the next hop neighbor.

Several problems with Watchdog have been identified in [23], such as the difficulty of unambiguously detecting that a node does not forward packets in the presence of collisions or in the case of limited transmission power. In addition to a watchdog-like observation, in CORE [10] nodes do not rely on promiscuous node. Rather, they can judge the outcome of a request by rating end-to-end connections. CONFIDANT [11] uses passive acknowledgement not only to verify whether a node forwards packets, but also as a means to detect if a packet has been illegitimately modified before being forwarded.

### 5.8 Response

Except for Watchdog and Pathrater [23], most of the reputation and trust systems have a punishment component for misbehaving nodes. The isolation of the misbehaving nodes is done in two steps: these nodes are avoided in routing and then denied cooperation when they request for it. Not using misbehaving nodes for routing but allowing them to use the network resources will only increase the incentive for misbehavior, since it results in power saving due to the decrease in number of packets they have to forward for others.

## 6. Examples of Reputation and Trust-based Models

In this section, various reputation and trust-based systems proposed in the literature for MANETs and WSNs are reviewed. For each of the schemes the working principle is discussed and critically analyzed in terms of its effectiveness and efficiency.

### 6.1 Watchdog and Pathrater

Watchdog and Pathrater components to mitigate routing misbehavior have been proposed by Marti, Giuli and Baker [23]. They observed increased throughput in MANETs by complementing DSR protocol with a *watchdog* for detection of denied packet forwarding and a *pathrater* for trust management and routing policy, rating every path used. This enables every node to avoid any malicious node in its routes as a reaction. Ratings are maintained for all the nodes in the network, and the ratings of actively used nodes are updated periodically. This scheme does not punish malicious nodes that do not cooperate in routing. Rather it relives the malicious nodes of the burden of forwarding for others, while their messages are forwarded without any problem. In this way, the malicious nodes are rewarded and reinforced in their behavior.

### 6.2 Context-Aware Inference Mechanism

A context-aware inference mechanism has been proposed by Paul and Westhoff [26] in which accusations are related to the context of a unique route discovery process and a stipulated time period. A combination is used that consists of unkeyed hash verification of routing messages and the detection of misbehavior by comparing a cached routing packet to overheard packets. The decision of how to trust nodes in future is based on accusation of others, whereby a number of accusations pointing to a single attack, the approximate knowledge of the topology, and context-aware inference are claimed to enable a node to rate an accused node with certainty. An accusation, however, has to come from several nodes. Otherwise if a single node makes an accusation, it is itself accused of misbehavior.

### 6.3 CORE

CORE (Collaborative Reputation Mechanism to enforce node cooperation in Mobile Ad hoc Networks) was proposed by Mirchiardi and Molva to enforce cooperation among nodes in MANETs based on a collaborative monitoring technique [10]. It differentiates between *subjective reputation* (observations), *indirect reputation* (positive reports by others), and *functional reputation* (task-specific behavior), which are suitably weighted to arrive at a *combined reputation* value. This combined reputation value is used to take decisions about cooperation or gradual isolation of a node. Reputation values are obtained by considering the nodes as *requestors* and *providers*, and comparing the expected result to the actually obtained result of a request. Essentially CORE is a distributed, symmetric reputation model that uses both first-hand and second-hand information for updating reputation. It uses bi-directional communication symmetry and dynamic source routing (DSR) protocol for routing. CORE also assumes wireless interfaces that support promiscuous mode of operation.

In CORE nodes have been modeled as members of a community and have to contribute on a continuing basis to remain trusted, else their reputation degrade, and eventually they are excluded from the network. The reputation is updated with time. As mentioned earlier, CORE uses three types of reputation: subjective reputation, indirect reputation, and functional reputation. CORE assigns more weight to the past observations than the current observations to ensure that a more recent sporadic misbehavior of a node has minimum influence on the evaluation of its overall reputation value. CORE has two types of protocol entities, a *requestor* and a *provider*.

- Requester: It is a network entity that requests for the execution of a function *f*. A requestor may have one or more providers within its transmission range.
- Provider: It is a network entity that can correctly execute the function *f*.

In CORE, nodes store the reputation values in a *reputation table* (RT), with one RT for each function. Each entry in the RT corresponds to a node and consists of four fields: (i) unique ID, (ii) recent subjective reputation, (iii) recent indirect reputation, and (iv) composite reputation for a predefined function. Each node is also equipped with a watchdog mechanism for promiscuous observation. RTs are updated during the request phase and the reply phase.

*Information Gathering:* The reputation of a node computed from first-hand information is referred to as subjective reputation. It is calculated directly from a node's observation. CORE does not differentiate between interactions and observations for subjective reputation unlike CONFIDANT [11]. The subjective reputation is computed only for the neighbors of the subject node. The subjective reputation is updated only during the request phase. If a provider does not cooperate with a requestor's request, then a negative value is assigned to the rating factor of that observation. This automatically decreases the reputation of the provider. The reputation of a node can take any value between –1 and +1. When a node joins the network for the first time, its reputation is initialized with a value 0.

*Information Sharing:* CORE uses indirect reputation, i.e., second-hand information to model MANETs. The impression of one node about another is influenced by other nodes in the network. However, there is a restriction on the type of reputation-information that can be propagated -only positive information exchange is allowed. As discussed earlier, this prevents bad mouthing attacks on benign nodes. Each reply message includes a list of nodes that cooperated in routing, and thus indirect reputation is updated only during the reply phase.

*Information Modeling:* CORE uses functional reputation to evaluate the trustworthiness of a node with respect to different functions. Functional reputation is computed by combining subjective and indirect reputation for different functions. Different applications may assign different weights to routing and various other functions like packet forwarding, etc. The combined reputation value of each node is computed by combining the three types of reputation with suitable weights. The positive reputation values are decremented with time to ensure that nodes cooperate and contribute on a continuous basis. This prevents a node from initially building up a very good reputation by being very cooperative and contributive but start misbehaving after some time.

*Decision Making:* When a node has to make a decision on whether or not to execute a function for a requestor, it checks the reputation value of the latter. If the reputation value is positive, the function is executed. However, the node is denied any service if its reputation is negative. A misbehaving node with low reputation value can build its reputation by cooperating with other nodes. However, reputation is difficult to build as it gets decreased every time the watchdog detects a non-cooperative behavior and also with time to prevent a malicious node from building reputation and then attacking the system resources.

*Discussions:* Assignment of more weight to the past reputation allows a malicious node to misbehave for some time if it has accumulated a high reputation value. False accusation attacks are prevented since only positive information is shared for indirect reputation updates. However, this makes the system vulnerable to false praise attack. The authors argue that a misbehaving node has no advantage by giving false praise to other unknown entities. This is true only so long as malicious nodes are not colluding. When malicious nodes start collaborating, then they can help prolong the survival time of another through false praise. However, the effect of false praise is mitigated in CORE to some extent by coupling the information dissemination to reply messages. Moreover, since only positive information is shared, the possibility of retaliation is prevented.

There is an inherent problem in combining the reputation values for various functions into a single global value. This potentially helps a malicious node to hide its misbehavior with respect to certain functions while behaving cooperatively with respect to other functions. The objective of a node to misbehave with respect to a particular function is to save its scarce resources. The node may choose to not cooperate for functions that consume resources like memory and power and choose to cooperate for functions that don't require these resources much. Nonetheless, functional reputation is a very nice feature of CORE that can be used to exclude nodes from functions for which their reputation value is below the threshold and include them for functions for which they have high reputation values. CORE also ensures that disadvantaged nodes that are inherently selfish due to their critical energy conditions are not excluded from the network using the same criteria as for malicious nodes. Hence, an accurate evaluation of the reputation value is performed that is not affected by sporadic misbehavior. Therefore, CORE minimizes false detection of the misbehavior of a node.

### 6.4 CONFIDANT

CONFIDANT (Cooperation Of Nodes-Fairness In Dynamic Ad-hoc NeTworks) is a security model proposed by Buchegger and Boudec [11] to make misbehavior unattractive in MANETs based on selective altruism and utilitarianism. It is a distributed, symmetric reputation model that uses both first-hand and second-hand information for computation of reputation values. CONFIDANT uses DSR protocol for routing and assumes that promiscuous mode of operation is possible. It does not require any tamper-proof hardware, since a malicious node neither knows its reputation values in other nodes nor does it have any access to those entries. The misbehaving nodes are punished by isolating them from accessing the network resources. Moreover, when a node encounters a misbehaving node, it sends a warning message to its trusted members in the network, termed as *friends*. CONFIDANT is based on the principle that reciprocal altruism is beneficial for every ecological system when favors are returned simultaneously because of instant gratification [36]. There may not be any benefit in behaving well if there is a delay in granting a favor and getting back the repayment. Each node runs four components in CONFIDANT protocol: *monitor*, *trust manager*, *reputation system*, and *path manager*.

*Information gathering:* The monitor module in each node passively observes the activities of the nodes within its 1-hop neighborhood. The node can detect any possible deviation made by the next node on the source route. It can also check for any possible content modification of the packets done by its next hop node. The monitor registers these deviations from normal behavior as soon as a bad behavior is detected, and reports this to the reputation system and the trust manager for evaluation of the new reputation value of the misbehaving node.

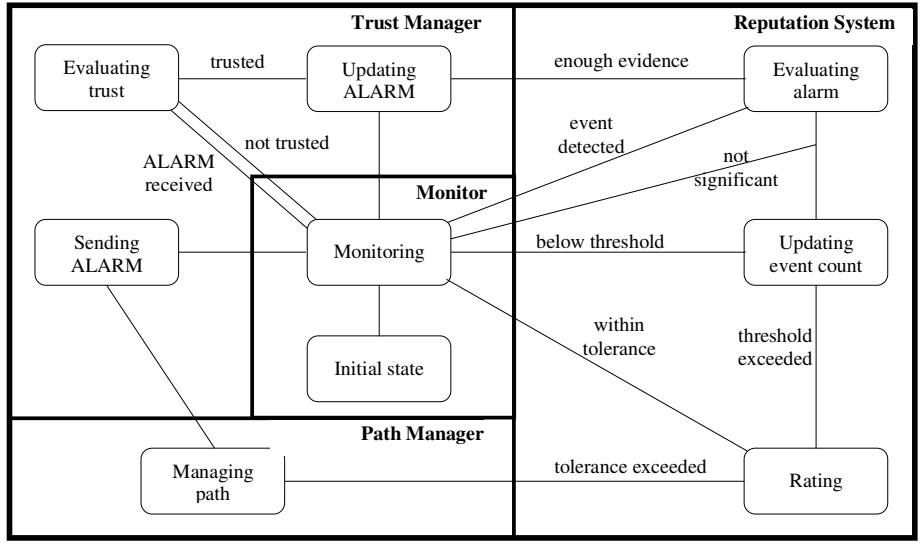

**Figure 2: Components and State Diagram of CONFIDANT Protocol [11]**

*Information Sharing:* The trust manager handles all the incoming and outgoing ALARM messages. Incoming ALARMs can originate from any node. Therefore, the source of an ALARM has to be checked for trustworthiness before triggering a reaction. This decision is made by looking at the trust level of the reporting node. CONFIDANT has provisions for several partially trusted nodes to send ALARMs which will be considered as an ALARM for a single fully trusted node. The outgoing ALARMs are generated by the node itself after having experienced, observed, or received a report of malicious behavior. The recipients of these ALARM messages are called friends, which are maintained in a friends list by each node.

The trust manager consists of three components: *alarm table*, *trust table*, and *friend list*. The alarm table contains information about received alarms, the trust table maintains the trust records of each node to determine the trustworthiness of an incoming alarm, and the friend list contains the list of all nodes to which the node has to send alarms when it detects any malicious activity. The trust manager is also responsible for providing with and receiving routing related information from other nodes in the network.

*Information Modeling:* The reputation system of every node maintains a table that consists of entries of other nodes and their corresponding reputation values. The reputation rating of a node is updated only when there is sufficient evidence of malicious behavior of that node occurring at least for a threshold number of times. The rating is changed using a function that assigns the highest weight on personal experience, a lesser weight for observations in the neighborhood and an even lesser weight to reported experience. The rationale behind this relative weighting scheme is that nodes trust their own experiences and observations more than those of other nodes. If the computed reputation value of a node falls below a predetermined threshold, the path manager is summoned for further actions.

*Decision Making:* The Path Manager is the component that is the decision maker. It is responsible for path re-ranking according to the security metric. It deletes paths containing misbehaving nodes and is also responsible for taking necessary actions upon receiving a request for a route from a misbehaving node.

*Discussions:* In CONFIDANT only negative information is exchanged between nodes. The authors argue that it is justified since malicious behavior is an exception and not the normal behavior. However, the exchange of only negative information makes the system vulnerable to false accusation attack on benign nodes by malicious nodes. Unlike CORE, even without collusion, malicious nodes benefit by falsely accusing benign node. With collusion of malicious nodes, this problem may become unmanageable. However, false praise attacks are possible since no positive information is exchanged. This prevents any possibility of collusion among a set of malicious nodes to prolong their survival time in the network. Since negative information is shared among the nodes, an adversary gets to know his situation and accordingly change his strategy. This may not be desirable.

Sharing negative information in the open may also introduce fear of retaliation that may force nodes to conceal their true findings.

In CONFIDANT, in spite of an elegant design of a reputation system, the reputation computation process using experienced, observed and reported information is not adequately explained. The nodes that are excluded because of misbehavior are allowed to recover after a certain timeout. This allows a malicious node to re-enter the system and attack repeatedly unless it is permanently denied entry after a certain number of times. Faulty nodes are treated in the same way as malicious nodes. This may not be always advisable as punishment may make the status of a faulty node even worse. The authors have not provided any reason for differentiating first-hand information as personal experience and direct observation and assigning them different weights.

### 6.5 OCEAN

OCEAN (Observation-based Cooperation Enhancement in Ad hoc Networks) has been proposed by Bansal and Baker as an extension of DSR protocol. It consists of a monitoring system and a reputation system [25]. In contrast to other approaches, in OCEAN nodes rely only on their own observations to avoid vulnerabilities arising out of false accusations and second-hand reputation exchanges. OCEAN categorizes routing misbehavior into two types: *misleading* and *selfish*. If a node has participated in a route discovery process but later on does not forward data packets, it is considered to be misleading as it misleads other nodes to route packets through it On the other hand, if a node does not even participate in the route discovery, it is considered to be selfish. In order to detect and mitigate the misleading behavior of nodes, after a node forwards a packet to one of its neighboring nodes, it buffers the packet checksum and monitors if the neighbor attempts to forward the packet within a given time. Depending on the activity of the neighboring node, its reputation rating is updated. If the rating falls below a threshold, the neighbor node is added to a faulty list, which is appended to the route request message as a list of nodes to be avoided in routing. All packets originating from the nodes in the avoid list are rejected so that the faulty nodes cannot use network resources. A *timeout* is used to allow faulty nodes to rejoin the network in case they may be wrongly accused or start behaving in a better manner. Each node also has a mechanism of maintaining *chipcount* for each of its neighbors to mitigate selfish behavior. A neighbor node earns chips when it forwards a packet on behalf of the node, and loses chips when it asks the node to forward a packet. If the chipcount of a node falls below a threshold, packets coming from that neighbor node are dropped.

### 6.6 Improved CONFIDANT- Robust Reputation System

Buchegger and Boudec presented an improved version of CONFIDANT called "A Robust Reputation System" (RRS) [27]. RRS introduced a Bayesian framework with Beta distribution for updating reputation. In contrast to CONFIDANT, RSS uses both positive and negative reputation values in the second-hand information.

RRS uses two different metrics: reputation and trust. The reputation metric is used to classify the nodes as either normal or misbehaving, while the trust metric is used to classify the nodes as either trustworthy or untrustworthy. Whenever second-hand information is received from a node, the information is put under a deviation test. If the incoming reputation information does not deviate too much from the receiving node's opinion, then the information is accepted and integrated with the current reputation value. Since the information sent by the reporting node is supported by the information previously maintained by the receiving node, the reporting node's trust rating is increased. On the other hand, if the reputation report deviates from the record maintained by the receiving node by more than a threshold value, then the reporting node's trust value is decreased. The receiving node also decides whether to integrate the deviating information with its current records, depending on the level of trustworthiness of the reporting node.

In RRS, only fresh information is exchanged. Unlike CORE, RRS gives more weight to current behavior than past. The authors argue that, if more weight is given to past behavior, then a malicious node can choose to be good initially till it builds a high reputation and trust value and then choose to misbehave. By assigning more weight to current behavior, the malicious node is forced to cooperate on a continuing basis to survive in the network.

### 6.7 RFSN

RFSN (Reputation-based Framework for Sensor Networks) is a distributed, symmetric, reputation-based model foe sensor networks, proposed by Ganeriwal and Srivastava [12]. It uses both first-hand and second-hand information for updating reputation values. In RFSN, nodes maintain the reputation and trust values only for their neighboring nodes, because there exist no sensor network applications in which a node requires prior reputation knowledge about a node that is multiple hops away. RFSN is the first reputation and trust-based model designed and developed exclusively for sensor networks. RFSN distinguishes between trust and reputation and uses two different metrics for their computation.

As shown in Figure 2, the first-hand information from the watchdog mechanism and second-hand information are combined to get the reputation value of a node. The trust level of the node is then computed from its reputation value. Based on this computed trust value, the node's strategy for the other node is determined. If the trust value is above a certain threshold, then the strategy is to cooperate with the node otherwise not.

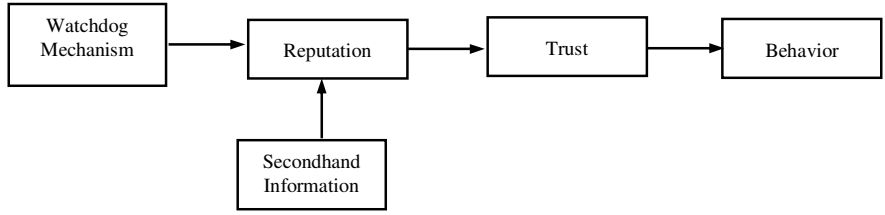

**Figure 2: Architecture of RFSN system [12]**

*Information Gathering:* RFSN, like many other systems, employs a watchdog mechanism for collecting first-hand information. The watchdog mechanism consists of different modules, each module monitoring a different function. The higher the number of modules, the greater is the resource requirement on the node. The reputation function is assumed to follow a probability distribution. The authors argue that reputation can only be used to statistically predict the future behavior of the nodes and it cannot be used to deterministically define the action performed by them. The reputation of all nodes that a node *i* interacts with is maintained in a reputation table in the node *i*. The direct reputation, $(R_{ij})_D$ is updated using the direct observations, i.e., the output of the watchdog mechanism.

*Information Sharing:* The nodes share their findings with each other. However, only positive information is shared. Higher weights are assigned to second-hand information from nodes that have higher reputation value associated with them. The weight assigned by node *i* to a second-hand information received from a node *k* is a function of the reputation of node *k* as maintained by node *i*. Like many other reputation and trust-based systems, RFSN uses Beta distribution model for reputation computation.

*Information Modeling:* Let us consider a node *i* that has established some reputation value, $R_{ij}$, for node *j*. Let the two nodes have $r + s$ further interactions among them, with *r* cooperative and *s* non-cooperative interactions respectively. The reputation value is now updated as follows:

$$R_{ij} = Beta\ (\ \alpha_j^{new} + 1, \beta_j^{new} + 1\ ) \qquad (1)$$

$$\alpha_j^{new} = (\ W_{age} * \alpha_j\ ) + r \qquad (2)$$

$$\beta_j^{new} = (\ W_{age} * \beta_j^{new}\ ) + s \qquad (3)$$

RFSN gives more weight to recent observations. This is done by using the factor $W_{age}$, which may take any value within the interval (0, 1). This is used for updating reputation value using direct observation. To update the reputation value using second-hand information, Dempster-Shafer theory [28] and belief discounting theory [29] are utilized. The reputation of a reporting node is automatically taken into account in the computation of the reputation of the reported node. This eliminates the need of a separate deviation test. A node with higher reputation gets a higher weight. The trust level of a node is determined using its reputation value. Trust is computed as the statistically expected value of reputation.

*Decision Making:* In the final decision making stage, a node *i* has to take a decision whether to cooperate with node *j*. The decision of node *i* is referred to as its behavior $B_{ij}$ and has a binary value: {*cooperate*, *don't cooperate*}. Node *i* uses the value of $T_{ij}$ to take the decision as follows:

$$B_{ij} = \begin{cases} cooperate & \forall\ T_{ij} \geq B_{ij} \\ don't\ cooperate & \forall\ T_{ij} < B_{ij} \end{cases} \qquad (4)$$

*Discussions:* RFSN treats misbehaving and faulty nodes the same way. The rationale is that a node that is uncooperative is to be excluded irrespective of the reason of its behavior. The nodes are allowed to exchange only good reputation information and only direct reputation information is propagated. This prevents the information from looping back to the initiating node. However, this increases the memory overhead slightly since a separate data structure has to be maintained for direct reputation.

### 6.8 DRBTS

DRBTS (Distributed Reputation and trust-based Beacon Trust System) model has been proposed by Srinivasan, Teitelbaum and Wu to solve a special problem in location-beacon sensor networks [13]. It is a distributed model that makes use of both first-hand and second-hand information. Two types of nodes are considered in the model: beacon node (BN) and sensor node (SN). The model is symmetric from the perspective of the BNs but asymmetric from the SNs' perspective. This is because BNs are capable of determining their location, and must pass this information to the SNs. However, without the knowledge of its own locations, an SN has no way of telling if a BN is lying to it. DRBTS enables the SNs to exclude location information from any malicious BN on the fly by using a simple majority principle. This way, DRBTS addresses the malicious misbehavior of any BN.

*Information Gathering:* In DRBTS, information gathering is addressed from two different perspectives: the SN's perspective and the BN's perspective. From a BN's perspective, DRBTS uses a watchdog for neighborhood watch. When an SN sends a broadcast asking for location information, each BN will respond with its location and reputation values for each of its neighbors. The watchdog packet overhears the responses of the neighboring BNs. It then determines its location using the reported location of each BN in turn, and then compares the value against its true location. If the difference is within a certain margin of error, then the corresponding BN is considered benign, and its reputation increases. If the difference is greater than the margin of error, then that BN is considered malicious and its reputation is decreased. From an SN's perspective, there is no first-hand information gathered by it through direct observations. The SNs rely completely on the second-hand information passed to them from nearby BNs during the location request stage. DRBTS also includes a method by which BNs can send out location requests disguised as SNs, in case of low network activity. However, unlike CONFIDANT, DRBTS does not differentiate first-hand information into personal experience and direct observation.

*Information Sharing:* DRBTS does not make use of second-hand information to update the reputation of its neighboring nodes. However, information sharing is only with respect to BNs. SNs do not share any information since they do not collect any virtue of their own observation of their neighborhood. In DRBTS, nodes are allowed to share both positive and negative reputation information. This is allowed to ensure a quick learning time.

*Information Modeling:* Let BN $j$ responds to a SN's request. Then BN $i$, in the range of $j$ updates its reputation entry of $j$ using this direct observation as follows:

$$R_{ki}^{New} = \mu_1 * R_{ki}^{Current} + (1 - \mu_1) * \tau \qquad (5)$$

where $\tau = 1$ if the location was deemed to be truthful and $\tau = 0$ otherwise. $\mu_1$ is a weight factor.

To use second-hand information, assume $B_j$ is reporting about BN $k$ to BN $i$. Now BN $i$ first performs a deviation test to check if the information provided by BN $j$ is compatible.

$$| R_{ji}^{Current} - R_{ki}^{Current} | \leq d \qquad (6)$$

If the above test is positive, then information provided is considered to be compatible and the entry $R_{ik}$ is updated as follows.

$$R_{j,i}^{New} = \mu_2 * R_{j,i}^{Current} + (1 - \mu_2) * R_{k,i}^{Current} \qquad (7)$$

If the deviation test in equation (6) is negative, then j is considered to be lying and its reputation is updated as follows.

$$R_{j,k}^{New} = \mu_3 * R_{j,k}^{Current} \qquad (8)$$

Equation (8) ensures that lying nodes are punished so that such misbehavior can be discouraged.

*Decision Making:* Decisions are made from the sensor node's perspective. An SN, after sending out a location request waits until a predetermined timeout. A BN has to reply before the timeout with its location information and its reputation ratings for its neighbors. Then, the SN, using the reputation ratings of all the responding BNs, tabulates the number of positive and negative votes for each BN in its range. Finally, when the SN has to compute its location, it considers the location information only from BNs with positive votes greater than negative votes. The remaining location information is discarded.

*Discussions:* DRBTS addresses the malicious behavior of beacon nodes. This unique problem that this system solves, though very important to a specific branch of WSNs, is not encountered very frequently. However, the idea can be easily extended to other problem domains.

## 7. Open Problems

Research in the field of reputation and trust-based systems for MANETs and WSNs is still in its incubation phase. There are many open issues that need to be resolved. One such issue is the bootstrapping problem. It takes an appreciable time for most of the reputation and trust-based systems to build trust among the nodes in the network. Developing an effective and efficient solution to minimize this latency is a big challenge. Utilizing all the available information in the network helps in making a reputation and trust-based system more aware about system-wide events, but it also makes the system vulnerable to false report attacks. Moreover, in systems like CORE where nodes have to contribute cooperatively on a continual basis to survive in the network, periods and regions of low network activities pose new challenges. Ageing will deteriorate the reputation of even benign nodes since there may not be any interactions among some nodes for some time.

Another important problem that needs to be addressed is devising a suitable defense against an intelligent adversary strategy. A sophisticated and intelligent adversary may manifest his attack strategy is such a way that it may not be possible for a detection system to catch him. A game theoretic approach may be applied here to investigate effectiveness of a detection and response system to defend against and counter such attacks.

As has been seen in this paper, a system like CORE uses functional reputation to monitor the behavior of nodes with respect to different functions. However, CORE unifies the reputation of a node for various functions into a global reputation value of that node. This may not be very effective in a real world scenario, since it will allow an adversary to conceal his misbehavior in certain functions while behaving very well for other functions. To the best of our knowledge, no research work so far has investigated the possible benefits of using functional reputation values independently. It may be effective for a security system to isolate a node from network resources for a particular function if the node is detected to be misbehaving with respect to that function, rather than judging it with respect to other functions where it may be perfectly well behaving.

Finally, another challenge especially for reputation systems in MANETs, is development of a robust scheme that motivates the nodes to publish their ratings honestly. This is not very easy, as the nodes in a MANET do not belong to the same interest group. However, in WSNs, since the nodes usually belong to some overarching system, there is an inherent motivation for nodes to participate honestly in information exchange.

## 8. Conclusion

Reputation and trust are two very important tools that have been used to facilitate decision making in diverse field like marketing, commerce and communication technology. This paper has provided a detailed understanding of reputation and trust-based systems both from the perspective of wireless communication networks. Many aspects of reputation and trust-based systems including their goals, properties, initialization process, and classification are discussed. Various important issues of design of such systems for wireless communication networks are also presented. A comprehensive review of some important research works focusing on adapting reputation and trust-based systems for MANETs and WSNs along with a critical evaluation of their strength and weaknesses are also presented. Finally, some open problems that are being currently investigated in this domain are also discussed.

**References**

[1] H. Liang, Y. Xue, K. Laosenthakul, and S. J. Lloyd, "Information Systems and Health Care: Trust, Uncertainty, and Online Prescription Filling, *Communications of AIS,* Vol. 15, pp. 41-60, 2005.
[2] D. H. McKnight, V. Choudhury, and C. Kacmar, "Developing and Validating Trust Measures for e-Commerce: An Integrating Typology, *Information Systems Research,* Vol. 13, pp. 334-359, 2002.
[3] D. M. Rousseau, S. B. Sitkin, R. S. Burt, and C. Camerer, "Not So Different After All: A Cross-Discipline View of Trust, *Academy of Management Review,* Vol. 23, pp. 393-404, 1998.
[4] D. Gefen, E. Karahanna, and D. W. Straub, "Trust and TAM in Online Shopping: An Integrated Model, *MIS Quarterly,* Vol. 27, pp. 51-90, 2003.
[5] P. Dasgupta, "Trust as a Commodity. In *Trust,* D. G. Gamretta, Ed. New York: Basil Blackwell, 1988, pp. 49-72.
[6] S. Ba and P. A. Pavlou, "Evidence of the Effect of Trust Building Technology in Electronic Markets: Price Premiums and Buyer Behavior, MIS Quarterly, Vol. 26, pp. 243-268, 2002.
[7] R. J. Lewicki and B. B. Bunker, "Trust in Relationships: A Model of Trust Development and Decline, In *Conflict, Cooperation and Justice,* B. Z. Rubin, Ed. San Francisco: Jossey-Bass, 1995, pp. 133-173.
[8] D. H. McKnight, L. L. Cummings, and N. L. Chervany, "Initial Trust Formation in New Organization Relationships, *Academy of Management Review*, Vol. 23, pp. 473-490, 1998.
[9] P. Mirchiardi and R. Molva, "Simulation-based Analysis of Security Exposures in Mobile Ad Hoc Networks" In *Proceedings of the European Wireless Conference*, 2002.
[10] P. Michiardi and R. Molva, "CORE: A COllaborative REputation mechanism to enhance node cooperation in Mobile Ad Hoc Networks, *Communication and Multimedia Security,* September 2002.
[11] S. Buchegger and J-Y. Le Boudec, "Performance Analysis of the CONFIDANT Protocol (Cooperation Of Nodes- Fairness In Dynamic Ad-hoc NeTworks), In *Proceedings of MobiHoc 2002*, Lausanne, CH, June 2002.
[12] S. Ganeriwal and M. Srivastava, "Reputation-based Framework for High Integrity Sensor Networks, In *Proceedings of the 2$^{nd}$ ACM Workshop on Security of Ad Hoc and Sensor Networks (SAN '04),* October 2004, pp. 66-77.
[13] A. Srinivasan, J. Teitelbaum, and J. Wu, "DRBTS: Distributed Reputation-based Beacon Trust System, In *Proceedings of the 2$^{nd}$ IEEE International Symposium on Dependable, Autonomic and Secure Computing (DASC'06),* Indianapolis, USA, 2006.
[14] M. Blaze, J. Feigenbaum, J. Ioannidis, and A. Keromytis, "RFC2704- The KeyNote Trust Management System Version 2", 1999.
[15] N. Li, J. Mitchell, and W. Winsborough, "Design of a Role-based Trust Management Framework", In *Proceedings of the IEEE Symposium on Security and Privacy,* Oakland, 2002.
[16] P. Ning and K. Sun, "How to Misuse AODV: A Case Study of Insider Attacks against Mobile Ad Hoc Routing Protocols, In *Proceedings of the 4$^{th}$ Annual IEEE Information Assurance Workshop*, West Point, June 2003.
[17] Y. Zhang and W. Lee, "Intrusion Detection in Wireless Ad Hoc Networks", In *Proceedings of the 6$^{th}$ Annual International Conference on Mobile Computing and Networking (MMOBICOM 2000,)* pp 275-283, ACM Press, New York, USA, 2000.
[18] P. Resnick, R. Zeckhauser, E. Friedman, and K. Kuwabara, "Reputation System", *Communications of the ACM*, 43(12): 4548, 2000.
[19] C. Dellarocas, "Immunizing Online Reputation Reporting Systems against Unfair Ratings and Discriminatory Behavior", *In Proceedings of the ACM Conference on Electronic Commerce*, pp. 150-157, 2000.
[20] P. Resnick and R. Zeckhauser, "Trust Among Strangers in Internet Transactions: Empirical Analysis of eBays's Reputation System", *Working Paper for the NBER Workshop on Empirical Studies of Electronic Commerce*, 2001.
[21] K. Aberer and Z. Despotovic, "Managing Trust in a Peer-2-Peer Information System", In *Proceedings of the 9$^{th}$ International Conference on Information and Knowledge Management (CIKM 2001)*, 2001.
[22] A. Josang and R. Ismail, "The Beta Reputaion System", In *Proceedings of the 15$^{th}$ Bled Electronic Commerce Conference*, Bled, Slovenia, June 2002.
[23] S. Marti, T. J. Giuli, K. Lai, and M. Baker, "Mitigating Routing Misbehavior in Mobile Ad Hoc Networks", In *Proceedings of the 6$^{th}$ Annual International Conference on Mobile Computing and Networking (MobiCom 2000)* 2000.
[24] S. Buchegger and J.-Y. Le Boudec, "The Effect of Rumor Spreading in Reputation Systems in Mobile Ad Hoc Networks", In *Proceedings of Wiopt' 03*, Sofia- Antipolis, March 2003.

[25] S. Bansal and M. Baker, "Observation-based Cooperation Enforcement in Ad Hoc Networks", *Research Report cs.NI/0307012,* Stanford University, 2003.
[26] K. Paul and D. Westhoff, "Context Aware Inferencing to Rate a Selfish Node in DSR-Based Ad Hoc Networks", In *Proceedings of the IEEE Globecom Conference*, Taipeh, Taiwan, 2002.
[27] S. Buchegger and J.-Y. Le Boudec, "A Robust Reputation System for Peer-to-Peer and Mobile Ad Hoc Networks", In *Proceedings of P2Pecon 2004*, Harvard University, Cambridge MA, USA, June 2004.
[28] G, Shafer, "A Mathematical Theory of Evidence", Princeton University, 1976.
[29] A. Jsang, "A Logic for Uncertain Probabilities", *International Journal of Uncertainty, Fuziness and Knowledge-Based Systems*, 9(3): 279-311, June 2001.
[30] P. Kollock, "The Production of Trust in Online Markets", *Advances in Group Processes,* edited by E. J. Lawler, M. Macy, S. Thyne, and H. A. Wlaker, 16, 1999.
[31] G. Montenegro and C. Castelluccia, "Statistically Unique and Cryptographically Verifiable (SUCV) Identifiers and Addresses, In *Proceedings of NDSS'02*, February 2002.
[32] J. R. Douceur, "The Sybil Attack", In Proceedings of the IPTPS '02 Workshop, Cambridge, MA, USA, March 2002.
[33] J. Newsome, E. Shi, D. Song, and A. Perrig, "The Sybil attack in Sensor Networks: Analysis and Defenses", In *Proceedings of the International Symposium on Information Processing in Sensor Networks*, 2004.
[34] Y. L. Sun, W. Yu, Z. Han, and K, J, Ray Li, "Trust Modeling and Evaluation for Ad Hoc Networks, *Technical Report No: 20041017-21*, University of Rhode Island, October 2004.
[35] J. Jubin and J. D. Tornow, "The DARPA Packet radio Network Protocols", In *Proceedings of the IEEE Communications and Networks*, 75(1), pp. 21-32, January 1987.
[36] R. Dawkins, "The Selfish Gene", Oxford University Press, 1989 edition.
[37] Y. Hu and A, Perrig, "A Survey of Secure Wireless Ad Hoc Routing", *IEEE Security And Privacy*, pp. 28-39, 2004.
[38] C. Karlof and D. Wagner, "Secure Routing in Wireless Sensor Networks: Attacks and Countermeasures", In *Proceedings of the 1st IEEE International Workshop on Sensor Networks Protocols and Applications*, pp. 113-127, May 2003.